\documentclass[aps,prl,twocolumn,groupedaddress,showpacs]{revtex4}

\usepackage{graphicx}% Include figure files
\usepackage{dcolumn}% Align table columns on decimal point
\usepackage{bm}% bold math

\usepackage{epstopdf}
\def\v_op{ \hat{\mathbf v} }

\newcommand{\br}{{\bf r}}
\newcommand{\bk}{{\bf k}}

\newcommand{\Nr}{{\mathcal N}}
\newcommand{\Nj}{N}
\newcommand{\calE}{\mathcal{E}}

\begin{document}

\title{Finite-size correction in many-body electronic structure calculations}

\author
{Hendra Kwee, Shiwei Zhang and Henry Krakauer}

\affiliation
{Department of Physics, College of William \& Mary,
Williamsburg, VA 23187-8795}

\date{\today}

\begin{abstract}
  
  Finite-size (FS) effects are a major source of error in many-body (MB)
  electronic structure calculations of extended systems. A method is presented
  to correct for such errors. We show that MB FS effects can be effectively
  included in a modified local density approximation calculation. A
  parametrization for the FS exchange-correlation functional is obtained.  The
  method is simple and gives post-processing corrections that can be applied
  to any MB results. Applications to a model insulator (P$_2$ in a supercell),
  to semiconducting Si, and to metallic Na show that the method delivers
  greatly improved FS corrections.

\end{abstract}


\maketitle

Realistic many-body (MB) calculations for extended systems are needed to
accurately treat systems where the otherwise successful density functional
theory (DFT) approach fails. Examples range from strongly correlated
materials, such as high-temperature superconductors, to systems with moderate
correlation, for instance where accurate treatments of bond-stretching or
bond-breaking are required.  DFT or Hartree Fock (HF), which are effectively
independent-particle methods, routinely exploit Bloch's theorem in
calculations for extended systems. In crystalline materials, the cost of the
calculations depends only on the number of atoms in the periodic cell, and the
macroscopic limit is achieved by a quadrature in the Brillouin zone, using a
finite number of ${\mathbf k}$-points. MB methods, by contrast, cannot avail
themselves of this simplification. Instead calculations must be performed
using increasingly larger simulation cells (supercells). Because the Coulomb
interactions are long-ranged, finite-size (FS) effects tend to persist to
large system sizes, making reliable extrapolations impractical. The resulting
FS errors in state-of-the-art MB quantum simulations often can be more
significant than statistical and other systematic errors. Reducing FS errors
is thus a key to broader applications of MB methods in real materials, and the
subject has drawn considerable attention \cite{PhysRevB.59.1917,ChiesaPRL06}.

In this paper, we introduce an external correction method, which is designed
to approximately include FS corrections in modified DFT calculations with {\em
  finite-size\/} functionals. The method is simple, and provides
post-processing corrections applicable to any previously obtained MB results.
Conceptually, it gives a consistent framework for relating FS effects in MB
and DFT calculations, which is important if the two methods are to be
seamlessly interfaced to bridge length scales. The correction method is
applied to a model insulator (P$_2$ in a supercell), to semiconducting bulk
Si, and to Na metal. We find that it consistently removes most of the FS
errors, leading to rapid convergence of the MB results to the infinite system.

We write the $\Nr$-electron MB Hamiltonian in a supercell as (Rydberg atomic
units are used throughout):
\begin{equation}
H= -\sum_{i=1}^\Nr\nabla_i^2 + \sum_{i=1}^\Nr V_{\mathrm{ion},i} + 
\sum_{i<j} V^{\rm FS}(|\br_i-\br_j|) \, ,
\label{eq:Hmb}
\end{equation}
where the ionic potential on $i$ can be local or non-local, and $\br_i$ is an
electron position. The Coulomb interaction $V^{\rm FS}$ between electrons
depends on the supercell size and shape, due to modification by the periodic
boundary conditions (PBC) \cite{PhysRevB.53.1814}. A FS correction is often
applied to the MB results from parallel DFT or HF calculations. The
corresponding DFT, as usually formulated, introduces a fictitious mean-field
$\Nr$-electron system \cite{PhysRev.136.B864,PhysRev.140.A1133}:
\begin{equation}
H_{\rm DFT} = -\nabla^2 + V_{\mathrm{ion}} + V_H(\br) +
V_{xc}^\infty(\br) \, , 
\label{eq:KS}
\end{equation}
where the Hartree and exchange-correlation (XC) potentials depend
self-consistently on the electronic density $n(\br)$. In the
non-spin-polarized local density approximation (LDA), for example:
$V_{xc}^\infty(\br)=\delta(n(\br)\,\epsilon_{xc}^\infty(n))/\delta n(\br)$,
where $\epsilon_{xc}^\infty(n)$ is typically obtained from quantum Monte Carlo
(QMC) results on the homogeneous electron gas (jellium), extrapolated to
infinite size \cite{PhysRevB.18.3126,PhysRevLett.45.566,PhysRevB.23.5048}.

Residual errors after DFT FS correction are still found to be large, however,
and the equations above illustrate why. The jellium QMC results, which
determine $\epsilon_{xc}^\infty(n)$, have been extrapolated to infinite
supercell size for each density. This is the correct choice for standard LDA
applications, where Bloch's theorem will be used to reach the infinite limit.
It is not ideal, however, if the LDA is expected to provide FS corrections.
Only one-body FS corrections (kinetic, Hartree, etc), which arise from
incomplete $\bk$-point integration, are included, while two-body FS
corrections \cite{PhysRevB.59.1917} are missing. If parallel HF calculations
are used instead, exact FS exchange $V_x \rightarrow V_x^{\rm FS}$ is
included, but $V_c$ is zero.

Our approach is to construct an LDA with FS XC in Eq.~(\ref{eq:KS}). If the
supercell of Eq.~(\ref{eq:Hmb}) is cubic (for simplicity), the XC energy is
$\epsilon^{\mathrm{FS}}_{xc}(n)\equiv\epsilon_x(r_s,L)+\epsilon_c(r_s,L)$,
where $r_s$ specifies the density via $4\pi r_s^3 /3 \equiv 1/n$ and $L$
denotes the linear size of the supercell. To obtain
$\epsilon^{\mathrm{FS}}_{xc}(r_s,L)$, we use unpolarized jellium systems in
the same supercell, in which the number of electrons $\Nj$ (distinct from
$\Nr$) is a variable \cite{FoulkesFSLDA2}, given by the ratio $r_s/L$:
$\Nj=(3/4\pi) (L/r_s)^3$.

\begin{figure}
\includegraphics[width=0.47\textwidth]{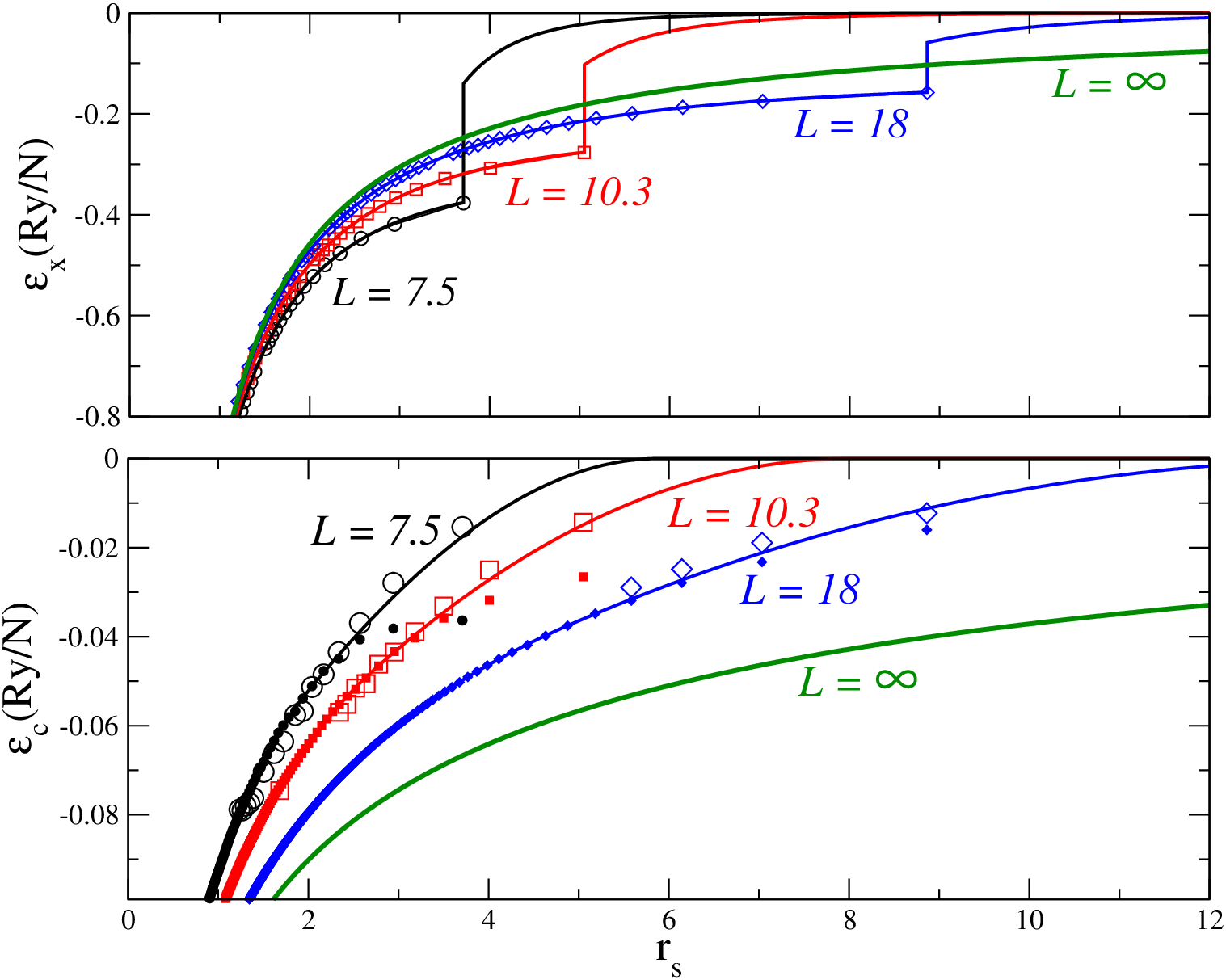}
\caption{(Color online) 
  Calculated and parametrized jellium exchange and correlation energies per
  electron vs.~$r_s$, for a range of supercell sizes $L$ (in Bohr). Top panel:
  exchange energies, with solid lines given by the fit in Eq.~(\ref{eq:Ex}),
  and open symbols by $\bk$-point averaged calculations.  Bottom panel:
  correlation energies, with solid lines given by the fit in
  Eq.~(\ref{eq:Ec}), open symbols by AF QMC calculations, and small filled
  symbols by the large-$N$ asymptotic expression in
  Eq.~(\ref{eq:extrapolation}).  }
\label{fig1}
\end{figure}

We parametrize the HF exchange energy in jellium by:
\begin{equation}
\epsilon_x(r_s,L)=
\cases{
      \frac{a_0}{r_s} + \frac{a_1}{L^2}r_s +  \frac{a_2}{L^3}r_s^2, 
      & if $r_s \le \gamma$;\cr
     \frac{a_3 L^5}{r_s^6},  & other.\cr
}
\label{eq:Ex}
\end{equation}
The term with $a_0 \simeq -0.916\,$Ry gives the usual infinite-size limit. A
$1/L$ canceling term, which arises from the self-interaction of an electron
with its periodic images \cite{PhysRevB.53.1814}, has been implicitly
included. The leading FS dependence is then $1/L^2$ \cite{PhysRevB.18.3126}.
The form of the remaining terms is motivated by the exact scaling relation:
$\epsilon_x(r_s,L)=\widetilde{\epsilon}_x(N)/L$. To obtain $\epsilon_x$, we
calculate $\widetilde{\epsilon}_x(\Nj)$ for a range of $\Nj$, each by
averaging over about 20 $\mathbf{k}$-points. The results are fitted to give
$a_1$ and $a_2$. As illustrated in Fig.~\ref{fig1}, the quality of the fit is
excellent. The behavior of $\epsilon_x$ at large $r_s$ requires special
handling for finite $L$. At $\gamma\equiv r_s(\Nj=2)$, there is only one
electron of each spin in the supercell, so $\epsilon_x$ is just the
self-interaction term. Beyond $\gamma$, $\epsilon_x$ is forced to go to zero
as $1/r_s^6$, reflecting the self-interaction of a `fractional' electron. The
coefficient $a_3$ is chosen to make the exchange potential $V_x^{\rm FS}$
continuous at $\gamma$ \cite{footnote1}. From Fig.~\ref{fig1}, the magnitude
of the discontinuity at $r_{s}=\gamma$ is seen to decrease with increasing
$L$, as expected. All parameters are listed in Table~\ref{tab:parameter}.

\begin{table}
\caption{\label{tab:parameter} Parameters (in Ry atomic units) 
in the FS XC functionals.
}
\begin{tabular}{lcccc}
\hline
\hline
$i$          &       1        &      2       &       3      &    4         \\
\hline
$a_{i}$ [Eq.~(\ref{eq:Ex})]   & $-2.2037$  & $0.4710$ & $-0.0150$ &  ---              \\
$g_{i}$ [Eq.~(\ref{eq:Ec})]   &   $0.1182$  & $1.1656$ & $-5.2884$ & $-1.1233$ \\
\hline
\hline
\end{tabular}
\end{table}

The correlation energy in jellium is the difference between the MB and HF
energies (per electron):
\begin{equation}
\epsilon_{c}(r_s,L)=\calE(r_s,L) -
t(r_{s},L) - \epsilon_x(r_s,L),
\label{eq:def_Ec}
\end{equation}
where the jellium non-interacting kinetic energy obeys the scaling relation
$t(r_s,L)=\widetilde{t}(\Nj)/L^2$. We calculate $\widetilde{t}(\Nj)$ in the
same way as $\epsilon_x(r_s,L)$, but averaging over more $\mathbf{k}$-points
to ensure convergence.

We next derive the MB energy $\calE$. Ceperley and Alder
\cite{PhysRevLett.45.566} obtained jellium QMC energies for various values of
$\Nj$ and provided the following fit:
\begin{equation}
\calE(r_s,L) = \calE^{\infty}(r_s)+b_1(r_s)\Delta t_\Nj +
b_2(r_s)/\Nj ,
\label{eq:extrapolation}
\end{equation}
where $\Nj$ uniquely determines $L$, and $\Delta t_\Nj=t(r_s,L)-t(r_s,\infty)$
is the FS error in the free-electron kinetic energy. The infinite-size limit,
$\calE^{\infty}$, was extrapolated from Eq.~(\ref{eq:extrapolation}) and it is
the basis for $V_{xc}^\infty$ in Eq.~(\ref{eq:KS}). The $b$ parameters were
given for several $r_s$ values, which we fit to get the functions $b_1(r_s)$
and $b_2(r_s)$. With these and $\Delta t_\Nj$, we can now calculate
$\calE(r_s,L)$ for any $r_s$ and $L$, which is accurate for large $\Nj$.

For small $\Nj$, namely large $r_s$ in a finite supercell,
Eq.~(\ref{eq:extrapolation}) does not apply.  This is easy to see from the
$1/\Nj$ term which, at sufficiently large $r_s$, causes $\epsilon_c$ to
diverge. To guide the analysis in this region, we use the plane-wave
auxiliary-field (AF) QMC method
\cite{PhysRevLett.90.136401,QMC-PW-Cherry:2007} to directly calculate $\calE$
for FS jellium systems. At small $r_s$, the correlation energy obtained is in
excellent agreement with that derived from Eq.~(\ref{eq:extrapolation}), as
shown in Fig.~\ref{fig1}. At large $r_s$, $\epsilon_c$ from
Eq.~(\ref{eq:extrapolation}) falls below the AF QMC value [$b_2(r_s)$ is
negative], as the latter goes to zero monotonically. The value of $r_s$ where
the two begin deviating depends on $L$, since it is determined by $N$.

We thus parametrize the correlation energy by
\begin{equation}
\epsilon_{c}(r_s,L)=\cases{
     \epsilon_c^{\infty}(r_s)-\frac{a_1}{L^2}r_s+\frac{g(r_s)}{L^3}, &
     $r_s \le \gamma_h$;\cr
      f(r_s), & $\gamma_h <r_s \le \gamma_l$;\cr
      0, &      other.\cr
}
\label{eq:Ec}
\end{equation}
The correlation functional has been divided into high, intermediate, and low
density regions. The boundaries are defined by $\gamma_h \equiv r_s(\Nj=12)$
and $ \gamma_l \equiv r_s(\Nj=1/2)$, which are guided by the discussion in the
previous paragraph and the quality of the fits described below, but are
otherwise arbitrary. At high densities, the infinite-size limit is given by
$\epsilon_c^{\infty}(r_s)$ (the Perdew-Zunger parametrization
\cite{PhysRevB.23.5048} is used here), and the leading FS term exactly cancels
that in $\epsilon_{x}(r_s,L),$ to ensure that $\epsilon_{xc}(r_s,L)$ correctly
scales as ${\mathcal O}(1/L^3)$. The function $g(r_s)\equiv g_1 r_s \ln(r_s) +
g_2 r_s + g_3 r_s^{3/2} + g_4 r_s^2$ is obtained from a fit to
$\epsilon_{c}(r_s,L)$ from Eqs.~(\ref{eq:def_Ec}) and
(\ref{eq:extrapolation}).  (The fits are illustrated in Fig.~\ref{fig1} and
parameters are given in Table~\ref{tab:parameter}.) At intermediate densities,
the function $f(r_s)$ is given by a cubic polynomial and is completely
determined by the requirement that $\epsilon_{c}$ and its derivative be
continuous at $r_s = \gamma_h$ and $r_s = \gamma_l$. As Fig.~\ref{fig1} shows,
the parametrization in Eq.~(\ref{eq:Ec}) closely reproduces our AF QMC data at
low densities for all cell sizes.

Post-processing FS corrections are now easily generated for any MB
calculation. The DFT$^{\mathrm{FS}}$ results, using
$\epsilon^{\mathrm{FS}}_{xc}$ from Eqs.~(\ref{eq:Ex}) and (\ref{eq:Ec}), can
be obtained from standard DFT computer codes with only minor modifications.
If $E^{\mathrm{FS}}(L)$ is the energy from DFT$^{\mathrm{FS}}$ and $E(L)$ from
standard DFT (i.e., DFT$^\infty$), the energy correction is
$\Delta\mathrm{DFT}^{\mathrm{FS}} =E(\infty)-E^{\mathrm{FS}}(L)$, where
$E(\infty)$ is obtained by $\bk$-point integration. The correction can
alternatively be expressed as the sum of $\Delta \mathrm{DFT}^{\mathrm{1B}}
\equiv E(\infty) - E(L) $ and $\Delta \mathrm{DFT}^{\mathrm{2B}} \equiv E(L) -
E^{\mathrm{FS}}(L)$. The one-body (1B) correction is the usual $\Delta
\mathrm{DFT}^{\infty}$, while the two-body (2B) part captures the FS effects
that arise from the modification of $V_{xc}$ due to supercell PBC.
 
\begin{figure}
\includegraphics[width=0.48\textwidth]{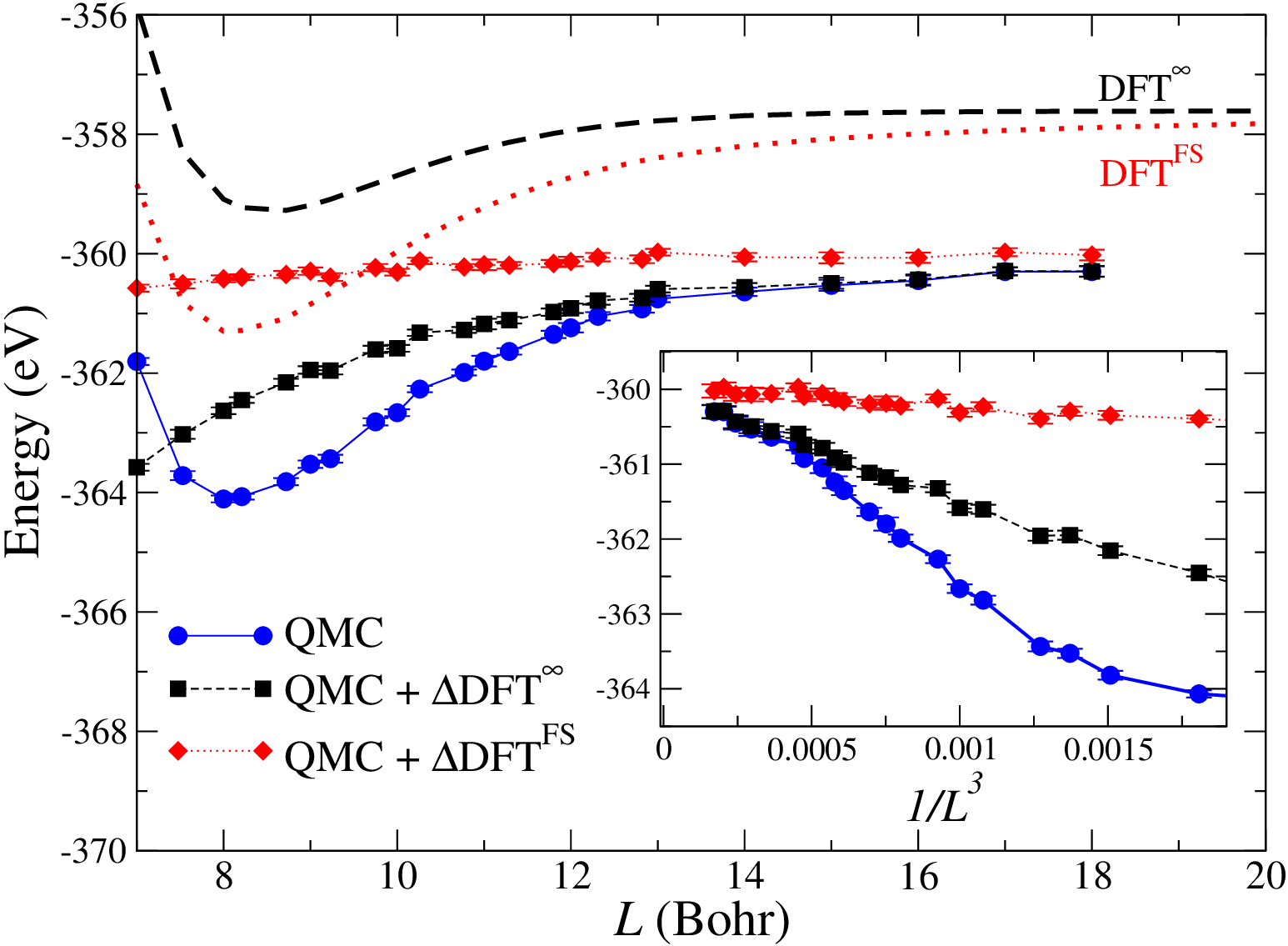}
\caption{(Color online) 
  $P_2$ PBC total energy convergence vs.~supercell size. Standard
  DFT$^{\infty}$ FS effect is different (too small) from that of MB AF QMC.
  DFT$^{\mathrm{FS}}$ parallels the MB calculation and leads to much more
  rapid convergence. The inset focuses on larger $L$ and shows the raw and
  corrected AF QMC results plotted as a function of $1/L^3$.}
\label{fig2}
\end{figure}

The present correction scheme is {\em exact\/} for homogeneous systems. Our
first application of the method is to a model system in the opposite limit. We
consider a ``molecular solid'' with P$_2$ in a periodic supercell, treated by
the plane-wave AF QMC method \cite{PhysRevLett.90.136401,QMC-PW-Cherry:2007}.
Because of the low-density ``vapor'' region and the variation in density, the
system provides a challenging test for the correction method. A
norm-conserving Kleinman-Bylander \cite{KB} separable non-local LDA
pseudopotential is used \cite{OPIUM}. Total energy calculations were performed
at the equilibrium bondlength of 3.578 Bohr, for cubic supercells of size
$L=7-18$ Bohr, all with the $\Gamma$-point ($\mathbf{k}=0$). Figure~\ref{fig2}
shows the results from AF QMC and LDA using both DFT$^{\infty}$ and
DFT$^{\mathrm{FS}}$ \cite{abinit}. The uncorrected QMC result has large FS
errors and, at $L=18$, is still $\sim$\,0.3\,eV away from the infinite-size
value. Corrected with DFT$^{\infty}$, the FS error is somewhat reduced at
intermediate $L$, but is unchanged for larger $L$ where the 2B effects
dominate. With the new method, the corrected energy shows excellent
convergence across the range, reaching the asymptotic value (within
statistical errors) by $L\sim 12$.

\begin{figure} 
\includegraphics[width=0.42\textwidth]{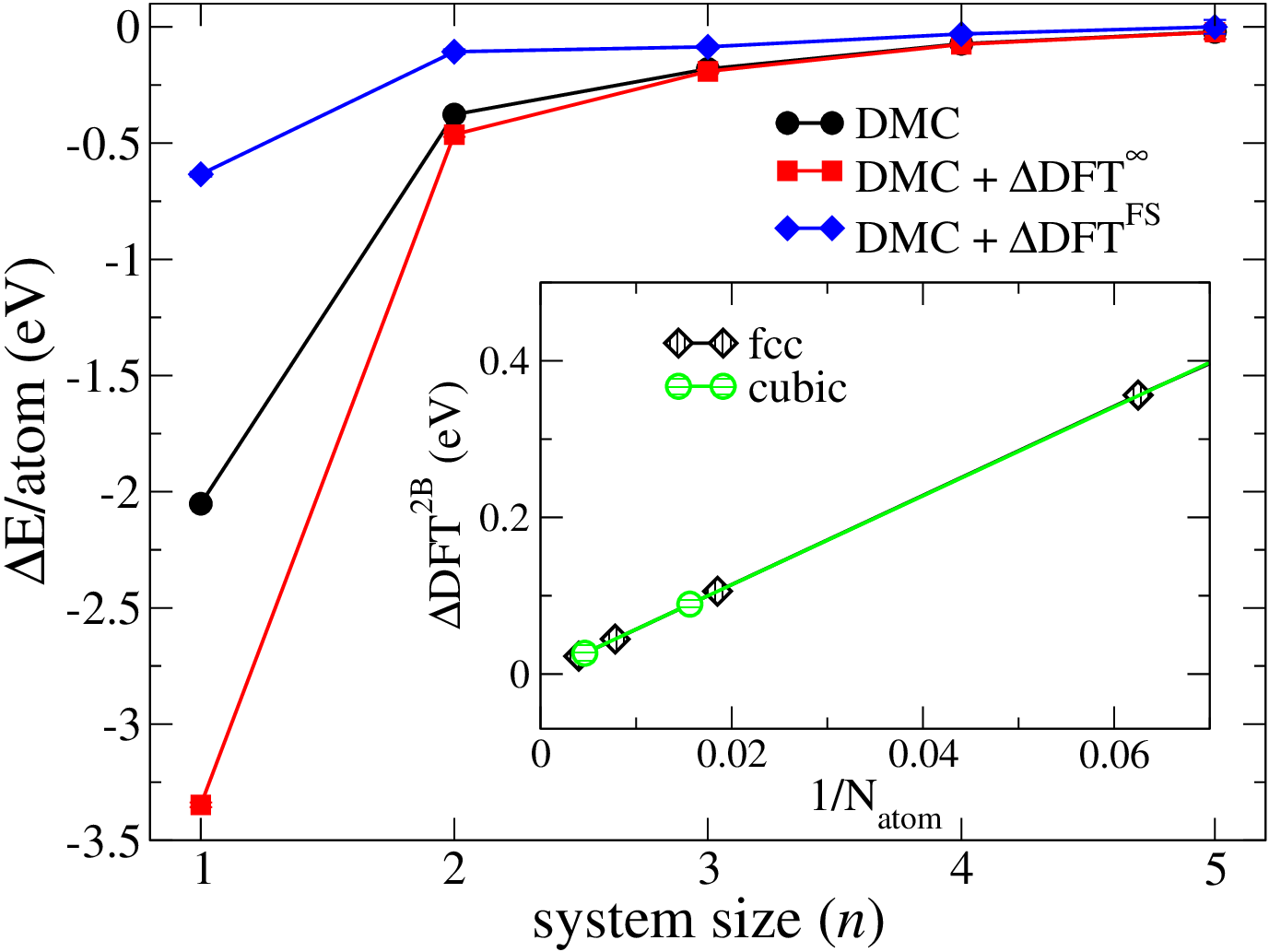}
\caption{(Color online) 
  Total energy per atom (in eV) of bulk Si in fcc supercells of size
  $n\!\times\!n\!\times\!n$ ($2n^3$ atoms). DMC energies in the main graph are
  from Kent {\it et.~al.\/} \cite{PhysRevB.59.1917} (shifted here relative to
  the extrapolated infinite-size limit). FS corrections are shown from both
  standard DFT and the present method. The inset shows the calculated two-body
  correction as a function of inverse the number of atoms, for fcc and cubic
  supercells.}
\label{fig3}
\end{figure}

The second application is for fcc bulk silicon, using {\em non-cubic\/}
supercells ($n\!\times\!n\!\times\!n$ the size of the primitive fcc cell). The
raw MB energies in Fig.~\ref{fig3} are taken from diffusion Monte Carlo (DMC)
calculations \cite{PhysRevB.59.1917}. In the FS corrections for the fcc
supercells, we use $\epsilon_{xc}^{\mathrm{FS}}$ from Eqs.~(\ref{eq:Ex}) and
(\ref{eq:Ec}), with an effective $L$ equal to the size of a {\em cubic\/}
supercell of the same volume.  The pseudopotential used is also different from
that in the DMC calculations. We checked multiple pseudopotentials to ensure
that the FS corrections are independent of the choice of pseudopotential. The
DMC calculations were done with the $\mathbf{k}=\mathrm{L}$ point. The usual
DFT correction is in the wrong direction in this case, thereby increasing the
FS error.  The new method removes most of the error, despite the non-optimal
$\epsilon_{xc}^{\mathrm{FS}}$. The inset in Fig.~\ref{fig3} shows
$\Delta\mathrm{DFT}^{\mathrm{2B}}$ calculated as above for fcc, compared with
that for cubic supercells. Both are seen to fall on an essentially smooth and
linear curve. This weak shape dependence of $\epsilon_{xc}^{\mathrm{FS}}$ is
encouraging, suggesting that additional FS MB jellium calculations can be
avoided in some non-cubic supercells.

\begin{figure}
\includegraphics[width=0.45\textwidth]{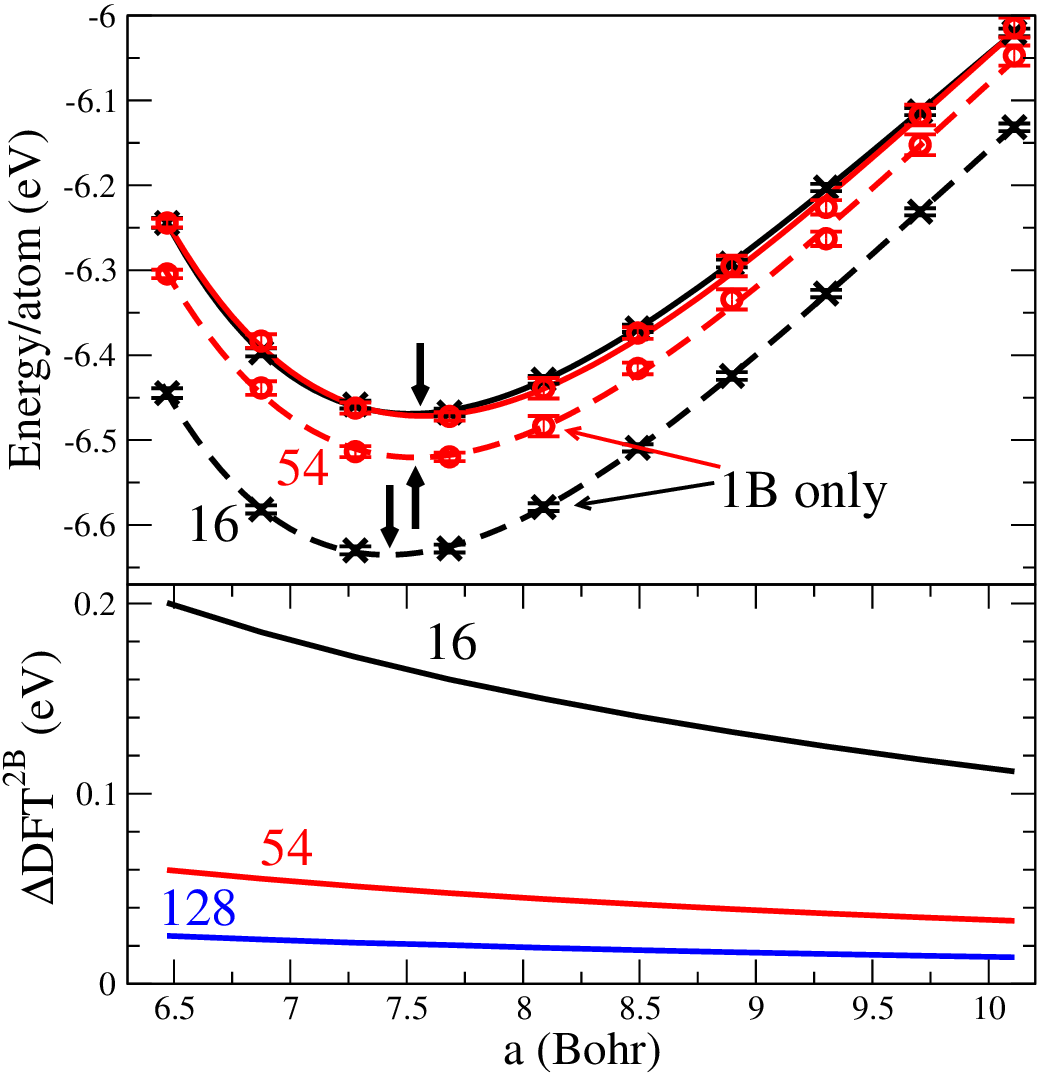}
\caption{(Color online) 
  Top: Equation of state for bcc Na. The AF QMC energy is shown vs.~lattice
  constant $a$ for 16- and 54-atom supercells, with one-body (1B) and full FS
  corrections. With full correction (solid lines), 16- and 54-atom results are
  almost indistinguishable. QMC statistical errors include that of ${\mathbf
    k}$-point averaging. The vertical arrows indicate the calculated
  equilibrium $a$.  Bottom: persistence of the two-body FS error, with finite
  slopes (number of atoms indicated).  }
\label{fig4}
\end{figure}

The final application is for metallic bcc bulk Na.  While in insulators a
single ${\mathbf k}$-point is often adequate, metals present additional
difficulties. We do multiple MB calculations with random ${\mathbf k}$-points
({\it e.g.} 50 for 16 atoms) and average the results
\cite{PhysRevE.64.016702}.  The plane-wave AF QMC method
\cite{PhysRevLett.90.136401,QMC-PW-Cherry:2007} was used, in which any
${\mathbf k}$-point can be included by a simple modification to the
one-particle basis. Although our pseudopotential has a Ne-core, DFT tests with
various pseudopotentials verified that it is sufficient for the cohesive
energy (consistent with Ref.~\cite{PhysRevB.68.165103}), but the frozen
semi-core introduces systematic biases in the lattice constant and bulk
modulus.
\begin{table}
\caption{\label{tab:cohesive} 
Calculated cohesive energy (in eV) of  bcc solid sodium vs.~experiment. 
AF QMC results from supercells with 2, 16 and 54 atoms are shown, 
together with FS-corrected results. A zero-point energy of 
$0.0145$\,eV/atom is included. Experimental value was taken 
from Ref.~\cite{PhysRevB.68.165103}.}
\begin{tabular}{llll}
\hline
\hline
     &                  &    \multicolumn{2}{c}{corrected} \\
      &     \ \  raw              &   w/ 1-body      &     w/ full FS    \\
\hline
$2$   &   $2.050(35)$ \ \quad  &   $2.141(2)$     &   $1.124(2)$ \\
$16$   &   $1.264(14)$   &   $1.287(4)$     &   $1.135(4)$ \\
$54$   &   $1.184(9)$    &   $1.189(10)$     &   $1.143(10)$ \\
expt  &                 &                  &   $1.129(6)$     \\
\hline
\hline
\end{tabular}
\end{table}
The calculated cohesive energies are given in Table~\ref{tab:cohesive}. The
FS-corrected cohesive energies for 16 and 54 atoms are consistent, and in
better agreement with experiment than the previous best DMC results
\cite{PhysRevB.68.165103} of $1.0221(3)$\,eV (with a core polarization
potential) and $0.9910(5)$\,eV (without). The calculated equation of state is
shown in Fig.~\ref{fig4}. We see that, with the new FS corrections and
${\mathbf k}$-point sampling, the calculations have better convergence than
previously reachable with an order of magnitude larger system sizes
\cite{PhysRevB.68.165103}.  Both the lattice constant and the bulk modulus
were modified by the FS corrections. As the bottom panel demonstrates, FS
effects always cause a {\em systematic\/} error in the lattice constant in
uncorrected MB calculations.

These tests show that our DFT$^{\rm FS}$ correction method works well in a
variety of systems. This is perhaps not surprising, given the often
near-sighted nature of the XC function. For the method to be effective, DFT
needs to provide a good approximation in capturing the {\em difference\/}
between the systems with interaction $V^{\rm FS}$ and $V$, which is not the
same as requiring DFT to work well in either system (assuming $L$ greater than
the size of the XC hole). We have presented an XC functional which delivers
high accuracy across several different materials. Previous attempts at FS
correction have focused on estimating the errors internally within the MB
simulation \cite{PhysRevB.59.1917,ChiesaPRL06}. Our approach is an external
method which is simple and can provide post-processing FS correction to any MB
electronic structure calculations. The method can be generalized, {\it
  e.g.\/}, to spin-polarized systems and other supercell shapes, and the FS
functional could be further improved, {\it e.g.\/}, by exact exchange.

We thank E.~J.~Walter for help with pseudopotentials, W.~Purwanto for help
with computing issues, and P.~Kent for sending us the numerical data from
Ref.~\cite{PhysRevB.59.1917}. This work is supported by ONR (N000140510055),
NSF (DMR-0535529), and ARO (48752PH) grants. Computing was done on NERSC and
CPD computers.


\begin{thebibliography}{17}
\bibitem[{foo()}]{footnote1}
\bibinfo{note}{In crystalline systems, the contribution from $r_s>\gamma$ is
  negligible for any reasonable size $L$ and we use a constant
  $\epsilon_x(\gamma,L)$ to make $\epsilon_x$ continuous.}
\end{thebibliography}
\end{document}